\documentstyle[preprint,aps]{revtex}

\newcommand{\pp}{\partial}
\newcommand{\w}{\mbox{\tiny $\wedge$}}
\newcommand{\spc}{\mbox{\hspace{1cm}}}

\begin{document}

\title{Higher dimensional Chern-Simons theories and topological black holes}

\author{M\'aximo Ba\~nados}

\address{Centro de Estudios Cient\'{\i}ficos de Santiago, Casilla 16443,
Santiago, Chile, and \\
Departamento de F\'{\i}sica, Universidad de
Santiago de Chile, Casilla 307, Santiago 2, Chile,}

\maketitle

\abstract{It has been recently pointed out that black holes of constant curvature with a ``chronological singularity" can be constructed in any spacetime dimension.  These black holes share many common properties with the 2+1 black hole. In this contribution we give a brief summary of these new black holes and consider them as solutions of a Chern-Simons gravity theory. We also provide a brief introduction to some aspects of higher dimensional Chern-Simons theories.}

\section{The topological black hole}

A topological black hole in $n$ dimensions can be constructed by making identifications along a particular Killing vector on $n$ dimensional anti-de Sitter space, just as the 2+1 black hole is constructed from 3 dimensional anti-de Sitter space. This procedure can be summarized as follows. Consider the $n$ dimensional anti-de Sitter space
\begin{equation}
-x_0^2 + x^2_1 + \cdots + x^2_{n-2} + x_{n-1}^2 - x_n^2 = -l^2,
\label{ads/n}
\end{equation}
and consider the boost $\xi =(r_+/l) (x_{n-1} \partial_{n} + x_n
\partial_{n-1})$ with
norm $\xi^2=(r_+^2/l^2)(-x_{n-1}^2+x_n^2)$. For $\xi^2=r_+^2$, one has the null surface,
\begin{equation}
x_0^2=x_1^2 + \cdots + x_{n-2}^2,
\label{h/n}
\end{equation}
while for $\xi^2=0$ one has the hyperboloid,
\begin{equation}
x_0^2=x_1^2 + \cdots + x_{n-2}^2 + l^2.
\label{s/n}
\end{equation}

Let us now identify points along the orbit of $\xi$. The region behind the hyperboloid ($\xi^2<0$) has to be removed from the physical spacetime because it contains closed timelike curves. The hyperboloid is thus a singularity because timelike geodesics end there.  On the other hand, the null surface (\ref{h/n}) acts as a horizon because any physical observer that crosses it cannot go back to the region which is connected to infinity. Indeed, the surface (\ref{h/n}) coincides with the boundary of the causal past of light like infinity. In this sense, the surface (\ref{ads/n}) with identified points represents a black hole. The existence of this $n$ dimensional anti-de Sitter black hole was first pointed out in \cite{Aminneborg+}, in four dimensions. 

Let us now introduce local coordinates on anti-de Sitter space (in the region $\xi^2>0$) adapted to
the Killing vector $\xi$. We introduce the
$n$ dimensionless local coordinates $(y_\alpha,\phi)$ by, 
\begin{eqnarray}
x_\alpha &=& \frac{2l y_\alpha}{1-y^2}, \mbox{\hspace{1cm} }
\alpha=0,...,n-2
\label{y} \nonumber\\
x_{n-1}  &=& \frac{lr}{r_+}  \sinh\left(\frac{r_+\phi}{l}\right),
\nonumber \\ 
    x_n  &=& \frac{lr}{r_+}  \cosh\left(\frac{r_+\phi}{l}\right),
    \nonumber 
\end{eqnarray}
with $r = r_+ (1+y^2)/(1-y^2)$ and $y^2 = \eta_{\alpha\beta}\, y^\alpha y^\beta $ [$\eta_{\alpha\beta}=\mbox{diag}(-1,1,...,1)$]. The
coordinate ranges are $-\infty < \phi < \infty$ and $-\infty < y^\alpha <\infty$ with the restriction $-1<y^2<1$. The induced metric has the Kruskal form,
\begin{equation} 
ds^2 =  \frac{l^2(r+r_+)^2}{r_+^2}\, dy^\alpha
dy^\beta\eta_{\alpha\beta} + r^2 d\phi^2, 
\label{ds/krus}
\end{equation}
and the Killing vector reads $\xi =\partial_\phi$ with $\xi^2=r^2$. In these coordinates, the quotient space is simply obtained by identifying $\phi \sim \phi+2\pi n$, and the resulting topology is $\Re^{n-1}\times S_1$. The metric (\ref{ds/krus}) represents the $n$-dimensional topological black hole black hole written in Kruskal coordinates. Note that the above metric is a natural generalization of the 2+1 black hole. Indeed, setting $n=3$ in (\ref{ds/krus}) gives the non-rotating 2+1 black hole metric written in Kruskal coordinates \cite{BHTZ}.

The fact that the metric (\ref{ds/krus}) is a natural generalization of the 2+1 black hole is of special relevance. Recently the 2+1 black hole has been shown to provide interesting applications in string theory \cite{new}.   

An important characteristic of the toplogical black hole for $n>3$ is the non-existence of a globally well defined timelike Killing vector. In other words, the black hole manifold is not static. This is easily seen by studying the behavior of the anti-de Sitter Killing vectors after the identifications are done \cite{Holst+}.  It is possible, however, to choose local static coordinates that resembles the Schwarszchild coordinates \cite{b3}.  

This problem does not appear when studying the ``Euclidean black hole". If we consider Euclidean anti-de Sitter space and then make identifications one can produce a non-trivial manifold that can be called the Euclidean topological black hole. Although in the Euclidean sector the notion of timelike does not make sense, one can identify a ``timelike" Killing vector $\pp_t$ which is globally defined.

By construction, the topological black hole has constant negative curvature and therefore it solves Einstein's equations with a negative cosmological constant.  However, for $n>3$, the ADM mass is infinity (see \cite{bgm} for the explicit calculations).  Here, we mean the ADM mass for the Euclidean black hole.  It was pointed out in \cite{b3} that finite conserved charges for the toplogical black hole can be defined in the context  of a Chern-Simons gravitational theory. In the next section we shall give a brief introduction to higher dimensional Chern-Simons theories and then consider the particular case of Chern-Simons gravity.

\section{Chern-Simons theory}

\subsection{The Lagrangian equations of motion}

A Chern-Simons theory can be regarded as a Yang-Mills theory with an exotic action. The main ingredients of a Chern-Simons theory in $n$ dimensions are; a $N$ dimensional Lie algebra with generators $T_a$ satisfying $[T_a,T_b] = f^c_{\ ab} T_c$ ($\ a=1,...,N$), a Yang-Mills gauge field $A  = A^a T_a$, and a $n-1$ totally symmetric invariant tensor $<T_{a_1},...,T_{a_{n-1}}> \equiv    g_{a_1 a_2 ... a_{n-1}}$. 

The Chern-Simons equations of motion can then be written as
\begin{equation}
g_{a_1 ... a_{n-1}} F^{a_2} \w F^{a_2} \w \cdots F^{a_{n-1}} = 0 
\label{cs/eq}
\end{equation}
where $F^a = dA^a + (1/2) f^a_{\ bc} A^b \w A^c$.  These equations of motion are derived from an action principle with a Lagrangian $L$ that satisfies $d L =g_{a_1 ... a_{n-1}} F^{a_1} \w \cdots \w F^{a_{n-1}}$. 

In the three dimensional case, these equations reduce to $g_{ab}F^b=0$. If $g_{ab}$ is non-degenerate, then they simply imply $F^a=0$ and the theory reduces to the problem of studying the class of flat connections modulo gauge transformations. The space of solutions of the equations of motion is thus completely classified given the topology of the manifold. 
An immediate consequence of this is that there are no local degrees of freedom in three dimensional Chern-Simons theory. 

In higher dimensions, however, the set of equations (\ref{cs/eq}) are far more complicated; they possess local degrees of freedom \cite{BGH1}  and the  space of solutions cannot be associated uniquely to the spacetime topology.  Furthermore, the phase space is stratified in regions with different number of degrees of freedom.  The maximum number of local physical degrees of freedom is equal to $mN -N - m$, with $n=2m+1$. This formula is valid only for $n>3$ and $N>1$. \footnote{For $N=1$ the Chern-Simons theory is surprinsingly more complicated because the separation of first nd second class constraints cannot be achieved in a generally covariant form. The $N=1$ theory in any number of dimensions does not have any local degrees of freedom.}       

Despite the complicated nature of the equations (\ref{cs/eq}), a full Hamiltonian decomposition can be performed \cite{BGH1}, and in this form the equations take a simple form.  Here we shall consider the equations of motion in a space+time form, in five dimensions.  The general situation as well as the Hamiltonian structure is analysed in \cite{BGH2}. 

\subsection{Space + time decomposition}

Suppose that locally we decompose the gauge field as 
\begin{equation}
A^a_\mu dx^\mu = A_0 dt + A_i dx^i,
\end{equation}
then the above equations, for $n=5$, can be split in the $4+1$ form
\begin{eqnarray}
\epsilon^{ijkl} g_{abc} 
F^{b}_{\ ij} F^{c}_{\ kl} &=& 0, \label{cons} \\
\Omega^{ij}_{ab} F^b_{j0} &=& 0, \label{dy} 
\end{eqnarray}
with
\begin{equation}
\Omega^{ij}_{ab} = \epsilon^{i j kl} g_{a b c} 
F^{c}_{\ kl}  .
\end{equation}
Note that contrary to the Yang-Mills equations of motion, these equations involve only the curvature tensor, and can be regarded as an algebraic set of equations for $F^a_{\mu\nu}$.  Thus, the integration of the higher dimensional Chern-Simons theory equations is equivalent to an algebraic problem plus solving the Bianchi identities $D F^a=0$. 

Equations (\ref{cons}) do not have any time derivatives and therefore they are constraints over the initial data. To our knowledge, the space of solutions of these equations is not known.  Equations (\ref{dy}), on the other hand,  do contain time derivatives. However, it is not obvious that there are no further constraints among them. 

The nature of the above equations is completely governed by the algebraic properties of the $4N\times 4N$  matrix $\Omega^{ab}_{ij}$, which depends on the invariant tensor $g_{abc}$ and $F^a_{ij}$.  Indeed, Eq. (\ref{dy}) depends explicitly on $\Omega$ while, using some simple combinatorial properties, Eq. (\ref{cons}) can also be written in terms of $\Omega$ as
\begin{equation}
\Omega^{ij}_{ab} F^b_{jk} = 0.
\end{equation}
In this form, the constraint is equivalent to the statement that $F^a_{ij}$ must be a zero eigenvalue of $\Omega^{ij}_{ab}$.  It has been shown in \cite{BGH1,BGH2} with many examples that generically there exists solutions to the constraint equations (\ref{cons}) for which the only zero eigenvectors of $\Omega$ are precisely $F^a_{ij}$. That is, if $V_i^a$ satisfies $\Omega^{ij}_{ab} V^b_i =0$, then there exists a vector field  $N^i$ such that $V^a_i = F^a_{ij} N^j $. The matrix $\Omega$ thus has four, and only four, zero eigenvalues. 

The space of solutions of the constraint satisfying this property carry the maximum number of degrees of freedom and we shall consider here only this sector of the theory. Note that, in particular, we exclude the flat solutions $F^a_{ij}=0$.    

We now turn to the dynamical equations. Equations (\ref{dy}) imply that $F^a_{i0}$ is a zero eigenvector of $\Omega^{ij}_{ab}$. The above discussion thus leads to the existence of a ``shift" vector $N^i$ such that
\begin{equation}
F^a_{i0} = F^a_{ij} N^j.
\end{equation}
Noting that $F^a_{i0} = \dot A^a_i - D_i A^a_0$ this equation is equivalent to the statement that the time evolution is generated by a gauge transformation with parameter $A^a_0$ plus a spatial diffeomorphism with parameter $N^i$.  The appearence of the spatial diffeomorphisms in the dynamical evolution reflects the fact that the gauge field is not flat and diffeomorphisms cannot be absorbed in the group of gauge transformations.

In the ``time gauge" $A^a_0=0$ and $N^i=0$, the dynamical equations simply imply $\dot A^a_i=0$ and therefore one is left only with the constraint equation (\ref{cons}). We shall see below that for the particular group $SO(4,2)$, the above equations of motion represent the  generalization of the Einstein equations in five dimensions due to Lovelock. Thus, if the constraint (\ref{cons}) was integrable, that would imply the integrability of the Chern-Simons Lovelock theory of gravity.  

\subsection{The $G \times U(1)$ theory}

We have seen in the last section that on the space of solutions for which $\Omega$ has the maximum rank, the equations of motion can be unambigously separated into constraints plus dynamical equations. However, we have not yet proved that the condition that $\Omega$ has only 4 null eigenvalues is not empty.  

The maximum rank condition can be explicitely implemented in a remarkably simple form if we couple to the original Chern-Simons action an Abelian $U(1)$ field, that we shall call $b$ \cite{BGH2}. As a matter of fact, we shall see below that an Abelian field with the correct coupling appears naturally in five dimensional Chern-Simons Supergravity. 

If we add to the original action the term $b\w F^a \w F^b g_{ab}$ with $g_{ab}$ the Killing form of the Lie algebra $G$, then the equations of motion are modified as
\begin{equation}
g_{abc} F^b \w F^c = H \w F^b g_{ab}, \spc F^a \w F^b g_{ab} =0,
\label{1}
\end{equation}
where $H = db$ is the field strength of the Abelian field.   These equations represent a Chern-Simons theory for the group $G\times U(1)$. Indeed, it is a simple exercise to prove that if we collect together the gauge field $A^A = (b,A^a)$, then there exists an invariant tensor $g_{ABC}$ of $G\times U(1)$ such that the above equations can be written as $g_{ABC} F^B \w F^C =0$.  

The usefulness of coupling the Abelian field $b$ is that now the maximum rank condition can be achieved simply by imposing that the pull back of $H$ to the spatial surface must be non-degenerate, that is det$(H_{ij}) \neq 0$.  To see this first note that the equations (\ref{1}) are solved by $F^a=0$ and $H$ arbitrary. Second, the matrix $\Omega$ evaluated on this particular solution has the block form
\begin{equation}
\left. \Omega^{ij}_{AB}\right|_{F^a=0} =  \left(\begin{array}{c|c}   
0_{4\mbox{x}4}   &  0_{4\mbox{x}4N}   \\ \hline
0_{4N\mbox{x}4}   &  g_{ab} \epsilon^{ijkl} H_{kl}     
\end{array}  \right).
\label{Ome} 
\end{equation}
The 4x4 zero block provides exactly the zero eigenvalues associated to the null eigenvectors $F^a_{ij}$ of $\Omega$. On the other hand, imposing $H_{ij}$ and $g_{ab}$ to be non-degenerate, the lower $4N$x$4N$ block is non-degenerate and $\Omega$ has indeed only four zero eigenvalues.  

We can now make perturbations with respect to this background. Since a non-zero eigenvalue cannot be set equal to zero by a small perturbation, 
the maximum rank condition is stable under small perturbations. 

\subsection{The $WZW_4$ algebra}

Perhaps the most interesting application of higher dimensional $G\times U(1)$ Chern-Simons theories is its relation with the $WZW_4$ theory proposed in \cite{Nair}, and further developed in \cite{Moore+}. These theories are generalizations of the standard two dimensional $WZW$ theories.

Let us first briefly review the relation between three dimensional Chern-Simons theory and the two dimensional $WZW$ model in the form developed in \cite{Moore-Seiberg}.  Due to the non-existence of local degrees of freedom in 3D Chern-Simons theory, one can solve the constraint $F^a_{ij}=0$ as $A_i = g^{-1} \pp_i g$, where $g$ is a map from the manifold to the group. Replacing back this value of $A_i$ into the Chern-Simons action one finds a Chiral $WZW$ action for the map $g$. The simplectic structure of the WZW model implies that the current $J(\lambda) = \int_{\pp \Sigma} Tr (\lambda A)$ satisfies the one dimensional Kac-Moody algebra \cite{Witten84}.  A different way to arrive at  the same result is by studying the issue of global charges \cite{Bimonti} in the Chern-Simons action. Indeed, if the Chern-Simons theory is formulated on a manifold with a boundary, then one can show that under appropriate boundary conditions, there exists an infinite set of global charges equal to $J(\lambda)$ that satisfy the Kac-Moody algebra. 

In the five dimensional $G\times U(1)$ Chern-Simons theory, the solution $F^a_{ij}=0$ to the constraint is by far not the most general one, although it is a good background in the sense that it carries the maximum number of degrees of freedom. Due to the existence of local degrees of freedom, one cannot solve the constraints in a close and general form, and therefore one does not find a simple model at the boundary. Still, one can study the issue of global charges and impose as a boundary condition that $A^a$ must be flat.   This has been done in detail in \cite{BGH2}. One finds an infinite tower of global charges given by $J(\lambda) = 
\int_{\pp \Sigma} H \w Tr (\lambda  A)$ and they satisfy the extension to three dimensions of the Kac-Moody algebra. 

\section{Five dimensional Chern-Simons Gravity}

\subsection{The action}

It is well known that in dimensions greater than four the Hilbert action is no longer the most general action for the gravitational field. For $D>4$, there exists a class of tensor densities that, as the Hilbert term, give rise to second order field equations for the metric, and a conserved energy momentum tensor\cite{Lovelock}. These terms are proportional to the dimensional continuation of the Euler characteristic of all dimensions $2p<D$ \cite{TZ}. 

For odd dimensional spacetimes there exists a particular combination of those terms such that the resulting theory can be regarded as a Chern-Simons theory of the form described in the last section\cite{Chamseddine}. Black holes solutions for this theory were found in \cite{BTZ3}.  

The simplest Chern-Simons theory of gravity exists in 2+1 dimensions with action
\begin{equation}
I_{2+1} = \int \epsilon_{abc} R^{ab} \w e^c.
\label{2m1}
\end{equation}
This action can be regarded as a Chern-Simons theory for the Poincare group. Indeed, besides local Lorentz rotations, (\ref{2m1}) is also invariant under Poincare translations,
\begin{equation}
\delta e^a = D \lambda^a, \spc  \delta w^{ab} = 0
\label{P}
\end{equation}
as can be easily verified using the Bianchi identity $D R^{ab}=0$.  
It is a simple exercise to prove that the 3+1 Hilbert counterpart of (\ref{2m1}) is not invariant under this transformation.   

The action (\ref{2m1}) can be extended to any odd dimensional spacetime.  For example, in five dimensions we consider
\begin{equation}
I_{4+1} = \int \epsilon_{abcde} R^{ab} \w R^{cd}\w  e^e.
\label{4m1}
\end{equation}
which is also invariant under (\ref{P}).  The key property that 
(\ref{2m1}) and (\ref{4m1}) share and makes them invariant under (\ref{P}) is that they are linear in the veilbein $e^a$. This is also the property that makes (\ref{4m1}) a Chern-Simons theory in five dimensions.  Just as (\ref{2m1}), the action (\ref{4m1}) has a simple supersymmetric extension \cite{BTrZ}.   

The actions (\ref{2m1}) and (\ref{4m1}) can be deformed to include a cosmological constant. For example, in five dimensions, the anti-de Sitter Chern-Simons theory of gravity is described by an action
\begin{equation}
I_{4+1}^{\Lambda} = \int \epsilon_{abcde} ( R^{ab} \w R^{cd}\w e^e + \frac{2}{3l^2} R^{ab}\w e^c\w e^d\w e^e + \frac{1}{5l^4} e^a\w e^b\w e^c\w e^d\w e^e)
\label{4m1l}
\end{equation}
where $l$ is a parameter with dimensions of length that parametrizes the cosmological constant. Note that apart from an overall constant (Newton's constant) and $l$, there are no other free parameters in this action. Apart from the explicit local Lorentz invariance, the action  (\ref{4m1l}) is also invariant under the deformed version of (\ref{P}),
\begin{equation}
\delta e^a = D\lambda ^a, \spc \delta w^{ab} = \frac{1}{l^2} (e^a \lambda^b - e^b \lambda^a)
\label{adst}
\end{equation}
that reduces to (\ref{P}) for $l^2\rightarrow\infty$.  As in the Poincare case, this action can be made supersymmetric (see \cite{Chamseddine,TrZ} and the contribution by J. Zanelli in this volumen) in a simple way. 

The transformations (\ref{adst}) plus the Lorentz rotations form a representation of the orthogonal group $SO(4,2)$. Although the action  is not explicitly invariant under this larger symmetry, the equations of motion following from  (\ref{4m1l}) can be collected as
\begin{equation}
\epsilon_{ABCDEF} \tilde R^{AB} \w \tilde  R^{CD} =0
\end{equation}
with $A=(a,6)$, $\tilde R^{ab} = R^{ab} + (1/l^2) e^a \w e^b $ and $\tilde R^{a6} = (1/l) T^a$. In this form, the $SO(4,2)$ symmetry is explicit.  

\subsection{Black holes}

An interesting property of the action (\ref{4m1l}) is the existence of two different black hole solutions for its equations of motion. One the one hand, there exists the topological black holes described in the first section with topology $\Re^4\times S_1$, constant curvature (zero anti-de Sitter curvature  $R^{AB}=0$), and a chronological singularity. On the other hand, the line element
\begin{equation}
ds^2 = - N^2 dt^2 + N^{-2} dr^2 + r^2 d\Omega_3 
\end{equation}
with 
\begin{equation}
N^2 = 1 - \sqrt{M+1} + \frac{r^2}{l^2} 
\end{equation}
is also an exact solution of (\ref{4m1l})  \cite{BTZ3}.  The constant $M$ is the ADM mass of the solution and one can see that an horizon exists only for $M>0$. For $M=-1$ one has anti-de Sitter space.  The scalar curvature of this metric is equal to 
\begin{equation}
R=-\frac{20}{l^2} + \frac{6 \sqrt{M+1}}{r^2}. 
\label{R}
\end{equation}
This geometry is thus singular at $r=0$ for all $M \neq -1$ and it approaches anti-de Sitter space asymptotically. This black hole has the topology $\Re^2 \times S_3$.

\subsection{Charges for the topological black hole}

As we saw in the last section, global conserved charges can be found in a simple form in a Chern-Simons theory provided one couples an Abelian gauge field adding a term to the action of the form $b\w g_{ab} F^a \w F^b$.  It turns out that in the context of five dimensional supergravity, this Abelian field is automatically present \cite{Chamseddine}. Indeed, supersymmetry requires an Abelian field $b$ coupled to the gravitational variables by the term $b \w R^{AB} \w R_{AB}$, where $R^{AB}$ is the anti-de Sitter curvature.   

We then consider the toplogical black holes as solutions to the Chern-Simons and compute their mass and angular momentum as explained above. [Angular momentum is added by using a different Killing vector to perform the identifications. See \cite{b3} and \cite{bgm} for more details.] The mass $M$ and angular momentum $J$ of the black hole embedded in this supergravity theory are,
\begin{equation} 
M = \frac{2r_+ r_-}{l^2}, \ \ \ \      
J = \frac{r_+^2 + r_-^2}{l}.
\label{J}
\end{equation}
In the same way one can associate a semiclassical entropy to the black hole which is given by 
\begin{equation}
S =  4\pi \, r_-.
\label{S}
\end{equation}      
This result is rather surprising because it does not give an entropy
proportional to the area of $S_1$ ($2\pi r_+$).  A similar phenomena has been reported by Carlip et al \cite{Carlip-Gegemberg-Mann}. The topological black hole thermodynamics in the context of standard general relativity has been analysed in \cite{Cr-Mann}.  

The entropy given in (\ref{S}) satisfies the first law, 
\begin{equation}
\delta M = T \delta S + \Omega \delta J, 
\label{fl}
\end{equation}
where $M$ and $J$ are given in (\ref{J}) and $T=(r_+^2-r_-^2)/(2\pi r_+l^2)$, $\Omega = r_-/lr_+$. 

During this work I have benefited from many discussions with Andy Gomberoff, Marc Henneaux, Cristi\'an Mart\'{\i}nez, Peter Peld\'an, Claudio Teitelboim, Ricardo Troncoso and Jorge Zanelli. This work was partially supported by the grant \# 1970150 from FONDECYT (Chile), and institutional support by a group of Chilean companies (Empresas Cmpc, Cge, Copec, Codelco, Minera La Escondida,
Novagas, Enersis, Business Design and Xerox Chile).

\end{document}